\documentclass{article}
 
\usepackage[english]{babel}
\usepackage{amsfonts,amsmath,dsfont}
\usepackage{amsthm}
\newcommand{\I}{\mathrm{i}}
\newcommand{\E}{\mathrm{e}}
\newcommand{\df}{\mathrm{d}}
\newcommand{\hsc}{h}

 \newcommand{ \tr }{ \mathrm{tr} }
  \newcommand{ \Tr }{ \mathrm{Tr} }

\newcommand{\hfast}{\mathcal{H}_{\mathrm{f}}}
\newcommand{\Or}{{\mathcal{O}}}
\newcommand{\R}{{\mathbb{R}}}

\newcommand{\D}{{\mathrm{d}}}
\newcommand{\Hi}{ \mathcal{H} }
\newcommand{\epsi}{ \varepsilon}

\theoremstyle{definition}

\theoremstyle{plain}

\theoremstyle{remark}

\title{Semiclassical approximations for   adiabatic slow-fast systems}

 \author{Stefan Teufel\\   Mathematisches Institut, Universit\"at T\"ubingen, Germany.}

\begin{document}

\maketitle

\begin{abstract}
In this letter we give a systematic derivation and justification of the  semiclassical model for the slow degrees
of freedom in adiabatic slow-fast systems first found by Littlejohn and Flynn~\cite{LF}. 
The classical Hamiltonian obtains a correction due to the variation of the adiabatic subspaces and the symplectic form is modified by the curvature of the Berry connection.
We show that this classical system  can be used to approximate quantum mechanical expectations and the time-evolution of operators   also in sub-leading order in the combined adiabatic and semiclassical limit.  
In solid state physics the corresponding semiclassical description of  Bloch electrons has led to substantial progress during the recent years, see \cite{Niu}. 
Here, as an illustration, we show how to compute the Piezo-current arising from a slow deformation of a crystal  in the presence of a constant magnetic field. 
  \end{abstract}

Consider   a quantum system   with a Hamiltonian $\hat H = H(x,-\I\epsi\nabla_x)$ given by the Weyl quantization of an  operator valued symbol $H(q,p)$  acting on a Hilbert space $\Hi= L^2(\R^n_x)\otimes  \hfast \cong L^2(\R^n_x, \hfast)$. Systems  composed of {\emph{slow}}   degrees of freedom with configuration space $\R^n_x$ and {\emph{fast}} degrees of freedom with state space $\hfast$ are of this form. The small dimensionless parameter $\epsi\ll 1$ controls the separation of time scales 
and the limit $\epsi\to 0$ corresponds to an adiabatic limit, in which   the slow and fast degrees of freedom decouple. At the same time  $\epsi\to 0$ is the  semiclassical limit for the slow degrees of freedom. Concrete realizations of this setting are the Born-Oppenheimer approximation \cite{LW}, the semiclassical limit of particles with spin \cite{LF2},   Bloch electrons in weak fields \cite{PST1} and many others, see also \cite{others} and references therein. 

In this letter we show that with each isolated eigenvalue $e (q,p)$ of the   symbol $H (q,p)$ there is   associated a classical system with an $\epsi$-dependent Hamilton  function $h(q,p) = e (q,p) + \epsi  M(q,p)$ and a modified symplectic form $
\omega(q,p) = \omega_0 + \epsi  \Omega(q,p)$. Here $M$ and $\Omega$ depend also on the corresponding spectral projection  $\pi_0(q,p)$ of $H(q,p)$. The correction to the energy $M$ results from a super-adiabatic approximation:   the true state of the fast degrees of freedom is in the range of  a slight modification $\hat \pi$  of the adiabatic projector $\hat\pi_0=\pi_0(x,-\I\epsi \nabla_x)$. 
The correction $\Omega$ to the symplectic form is given by the curvature of the Berry connection and takes into account the geometry of the eigenspace bundle defined by $\pi_0(q,p)$.
We show that with the help of this classical  system  one can approximate   expectations for ``slow'' observables $\hat a\otimes \mathbf{1}_{\hfast}$    and   also their time-evolution   in the Heisenberg picture   with errors of order~$\Or(\epsi^2)$.

The classical system  described  above appeared first  in the seminal  work of Littlejohn and Flynn~\cite{LF} in the context of WKB approximations. But they neither claimed nor proved the statements of the present paper. 
Independently, Niu and coworkers \cite{Niu} applied the  classical model   with enormous success in the context of the semiclassical description of Bloch electrons. Here the  classical description  provides simple and straightforward  derivations   of  formulas that are hard  to justify by other means. 

Given the abundance of slow-fast systems in physics,  a complete understanding of their semiclassical limit is desirable. 
The main novelty presented in this letter are general and systematic proofs showing that the classical model indeed approximates the quantum mechanical expectations of time-dependent ``slow'' observables. Our approach differs from earlier ones in two ways. Based on \cite{EW}, it is intrinsically gauge invariant as it does not use a local choice of eigenfunctions for the fast system. And it directly applies to arbitrary states, not only to semiclassical wave packets. For a mathematically rigorous formulation of our results we refer to \cite{StTe}.

 
The structure of this letter is as follows.
We first recall a few basic facts about  the Weyl calculus and the adiabatic approximation. Then we construct the classical Hamiltonian system and prove the various semiclassical approximations. Finally we show how to incorporate also time-dependent Hamiltonians and compute, as an illustration,  the Piezo-current in the presence of a magnetic field. We conclude with some remarks on the literature.

\section{Preliminaries and definitions} 

  \subsection{Weyl calculus}
 
 With a   function  $A  : \R^{2n}\to \mathcal{L}(\hfast)$ on classical phase space taking values in the linear operators on $\hfast$ one associates the Weyl operator  
 \[
(\hat{A}\psi)(x)=\frac{1}{(2\pi \epsi)^n}\int_{\mathbb{R}^{2n}}\hspace{-8pt}\df p \df y  A\left(\tfrac{1}{2}(x+y),p\right)\E^{\I p \cdot (x-y)/\epsi} \psi (y)
\]
acting on $\psi \in\Hi=L^2(\R^n_x, \hfast)$.
  The composition of operators $\hat A\hat B= \hat C$ induces a composition of symbols  denoted by
$C= A\# B$ and called the 
Moyal product. The asymptotic expansion of $A\# B$ starts with
\[
A\# B \asymp  A_0 B_0 + \epsi ( A_1 B_0 + A_0B_1 - \tfrac{\I}{2} \{ A_0, B_0\}
 )+\Or(\epsi^2),
\]
where $\{\cdot,\cdot\}$ denotes the Poisson bracket
$
\{ A_0, B_0\}   =   \sum_{j=1}^n  \left( \partial_{p_j} A_0 \, \partial_{q_j} B_0  - \partial_{q_j} A_0\, \partial_{p_j} B_0\right)
$. 
Since $A_0$ and $B_0$ are operator valued functions, they do not commute in general and neither do their derivatives. Hence, in general, $\{A,A\}\not=0$, but, if the derivatives of $A$ are trace-class,
\begin{equation}\label{cyctrace}
\tr \left(\{A,A\}\right) =0\,,
\end{equation}
because of the cyclicity of the trace. 
For later reference we state also the subprincipal symbol for triple products 
\begin{eqnarray}
\label{eq:triple}\lefteqn{\hspace{-5mm}
 (A\#B\#C)_1 =    A_1B_0C_0+A_0B_1C_0+A_0B_0C_1} \\ \nonumber
 && -\tfrac{\I }{2} ( A_0\{B_0,C_0\}+\{A_0,B_0\} C_0+ \{A_0|B_0|C_0\} ) \,.
\end{eqnarray}
Here and in the following we use the shorthand
\[
 \{A_0|B_0|C_0\}  \;:=\; \partial_p A_0\cdot B_0\,\partial_q C_0  - \partial_q A_0\cdot B_0 \,\partial_p C_0\,.
\]
If a symbol $A = a \cdot{\rm id}$ is a scalar multiple of the identity, then $A$ and all its derivatives commute with any $B$. As a consequence one can show that in this case
\begin{equation}\label{weylcommu}
A_0\# B_0 - B_0\# A_0 \asymp -\I\epsi  \{ A_0, B_0\} \,+\, \Or(\epsi^3)\,.
\end{equation}
The fact that the remainder term in (\ref{weylcommu}) is of order $\epsi^3$ and not only $\epsi^2$ is at the basis of   higher order semiclassical approximations. 
Note that $\pi_0^2=\pi_0$ implies that  any partial derivative  $\partial_j\pi_0$  is off-diagonal with respect to~$\pi_0$, 
\begin{equation}\label{pi0off}
\partial_j\pi_0 = \pi_0 ( \partial_j\pi_0)  \pi_0^\perp+  \pi_0^\perp (\partial_j\pi_0)\pi_0\,,
\end{equation} 
with   $\pi_0^\perp = 1-\pi_0$. Thus for scalar symbols $a= a\mathbf{1}_{\hfast}$  
 \begin{equation} 
 \label{eq:pb2}
  \{a,\pi_0\} = \pi_0 \{a,\pi_0\} \pi_0^\perp + \pi_0^\perp\{a,\pi_0\} \pi_0. 
\end{equation}
Finally one can express the trace of a product of Weyl operators by a classical phase space integral,
  \begin{equation}
\Tr \big(\hat A\hat B\big) = \frac{1}{(2\pi\epsi)^n} \int_{\R^{2n}}\hspace{-5pt}\D q\D p\;\tr (A(q,p) B(q,p)) \,.\label{traceform}
\end{equation}
 Here  $\Tr$ denotes the trace on $\Hi$ and $\tr$ the trace on $\hfast$.

\subsection{Adiabatic approximation}
Let $e_0(q,p)$ be a non-degenerate eigenvalue band of the principal symbol $H_0(q,p)$ of the Hamiltonian  $\hat H= \hat H_0 + \epsi \hat H_1$ and $\pi_0(q,p)$ the corresponding family of rank one  spectral projections. Then there exists an associated subspace $\hat\pi\Hi$ of $\Hi$ 
that is adiabatically invariant: \\[1mm] 
\noindent   {\bf Claim:} 
There exists a projection~$\hat \pi$ with symbol
$ \pi \; = \; \pi_0 + \epsi \pi_1 +\Or(\epsi^2)$
such that   $[\hat \pi, \hat H] =\Or(\epsi^\infty)$. As a consequence, the range of $\hat \pi$ is almost invariant, $[\hat \pi,  \E^{-\I \hat H \frac{t}{\epsi}} ]  =\Or(\epsi^\infty|t|)$,
and   the adiabatic approximation
$
 \E^{-\I \hat H \frac{t}{\epsi}} \hat \pi =  \E^{-\I\hat \pi \hat H\hat \pi \frac{t}{\epsi}} \hat \pi +\Or(\epsi^\infty|t|) 
$ holds.
Moreover
\begin{equation}\label{pi1d}
\pi_0 \pi_{1}\pi_0 =\tfrac{\I}{2}\,\pi_0 \{\pi_0,\pi_0\} \pi_0\,.
\end{equation}
\noindent {\em Proof.} The construction of $\pi$ was first done in \cite{EW} and is by now standard, c.f.\ \cite{others}. Equation (\ref{pi1d})  simply follows from the fact that $\pi$ is a projector in the Moyal algebra, 
$0 = \pi \# \pi - \pi =     \epsi( \pi_1\pi_0 + \pi_0\pi_1 -\tfrac{\I}{2}\{ \pi_0,\pi_0\} - \pi_1 
) +\Or(\epsi^2)$.

\subsection{The classical Hamiltonian}
It therefore suffices to study  the restriction $\hat\pi \hat H\hat\pi$ of $\hat H$ to the adiabatic subspace $\hat\pi\Hi$
in order to understand the dynamics within $\hat\pi\Hi$.
We will show that its semiclassical limit  is governed by the  associated   scalar Hamilton function 
\[
\hsc :=e+\epsi M,
\]
with $e=e_0+\epsi \tr(H_1\pi_0)$ and
$M:=\tfrac{\I}{2}\,\tr \left(\{\pi_0|H_0|\pi_0\}\right)$.\\[1mm]
{\bf Claim:} It holds that 
$\pi \# \hsc  \# \pi  - \pi \# H \# \pi = \Or(\epsi^2)$ 
and thus $\hat\pi \,\hat \hsc \,\hat\pi - \hat\pi \,\hat H \,\hat\pi  = \Or(\epsi^2)$.\\[1mm]
 \noindent {\em Proof:} In the expansion of  $\pi \# \hsc  \# \pi - {\pi}\#{H}\#{\pi}$ the principal symbol vanishes 
  and the subprincipal symbol is 
$M \pi_0     + \tfrac{\I}{2}(\pi_0\{H_0-e_0,\pi_0\}+  \{\pi_0,H_0-e_0\}\pi_0
+ \{\pi_0|H_0-e_0|\pi_0\})$.   
  To see that it also vanishes, note that (\ref{cyctrace}) and  (\ref{pi0off}) imply
\begin{eqnarray*}
M\pi_0 &=&\tfrac{\I}{2} \tr \left(\{\pi_0|H_0|\pi_0\}\right)\pi_0  
\stackrel{(\ref{cyctrace})}{=}   \tfrac{\I}{2} \tr \left(\{\pi_0|H_0-e_0|\pi_0\}\right)\pi_0\\
&\stackrel{(\ref{pi0off})}{=}& \tfrac{\I}{2} \tr \left(\pi_0 \{\pi_0|H_0-e_0|\pi_0\}\pi_0\right)\pi_0\\
&=& \tfrac{\I}{2} \pi_0 \{\pi_0|H_0-e_0|\pi_0\} \pi_0 
 \stackrel{(\ref{pi0off})}{=}   \tfrac{\I}{2}   \{\pi_0|H_0-e_0|\pi_0\} \,.
\end{eqnarray*}
Moreover
$
0 =   \partial_j \big( (H_0-e_0) \pi_0\big)\pi_0 =  \partial_j  (H_0-e_0)  \pi_0 +   (H_0-e_0) \partial_j  \pi_0\pi_0
$
implies that 
$
- \{\pi_0,H_0-e_0\}\pi_0 =  \{\pi_0| H_0-e_0|\pi_0\}\pi_0 =  \{\pi_0|H_0-e_0|\pi_0\},
$
which proves the claim and shows also 
$M = \tfrac{\I}{2} \tr \left(\{\pi_0|H_0-e_0|\pi_0\}\right)  =-\tfrac{\I}{2} \tr \left(\pi_0\{\pi_0,H_0-e_0\}\right)$.

\subsection{The symplectic form and its Liouville measure}
 
In order to obtain semiclassical approximations up to errors of order $\epsi^2$, one needs to take into account that 
the restriction to the range of $\hat \pi$ also induces a modified symplectic form $\omega_\epsi $ on $\R^{2n}$ given by 
\[
{\omega_\epsi}  := \omega_0 + \epsi \,\Omega =
 \left(\hspace{-3pt} \begin{array}{cc} 0 &  E_n\\  -E_n & 0\end{array}\hspace{-3pt}\right)\;+\;\epsi \left( \hspace{-3pt}\begin{array}{cc} \Omega^{qq} &  \Omega^{qp}\\  \Omega^{pq} & \Omega^{pp}\end{array}\hspace{-3pt}\right),
\]
where the
  components of $\Omega$ in the canonical basis are
  \[
  \Omega_{\alpha\beta} \;:=\; -\,\I \,\tr_{\hfast}\left(\pi_0[\partial_{z_\alpha}\pi_0,\partial_{z_\beta}\pi_0] \right) \,,
\]
with  $z=(q,p)$ and $\alpha,\beta =1,\ldots,2n$.
By definition $\Omega$ is skew-symmetric and one readily checks that it defines a closed 2-form. Actually  $\Omega$ is the curvature 2-form of the Berry connection.  For $\epsi$ small enough $\omega_\epsi$ is thus a symplectic form.
The  Liouville measure $\lambda_\epsi $ associated with the symplectic form $\omega_\epsi $ has the expansion 
\begin{eqnarray*}\label{liouville} 
 \lambda_\epsi&=&  \Big(1+\tfrac{\epsi}{2} \, \sum_{j=1}^n \left(\Omega^{qp}_{jj} -\Omega^{pq}_{jj} \right)\Big) \df q^1\wedge  \dots \wedge \df p^n  +\Or(\epsi^2)
 \nonumber\\
 &=& \left( 1+\I \epsi \tr\left(\pi_0\{\pi_0,\pi_0\}\right)\right) \df q^1\wedge\dots\wedge\df p^n+\Or(\epsi^2) . 
\end{eqnarray*}
 
\section{Results: Semiclassical approximations} 

\subsection{Equilibrium expectations}

Let $a:\R^{2n}\to\R$ be   integrable    and $f:\R\to \R$ be smooth, then
\begin{equation}\label{expt}
\Tr \left(\hat\pi f(\hat H)\, \hat a\right)  =  \frac{1}{(2\pi\epsi)^n}\int  \D \lambda_\epsi \; f(\hsc (q,p)) \, a(q,p)  + \Or(\epsi^{2-n}).
\end{equation}
  
\noindent {\em Proof:} In the following computation we use  that $[\hat \pi,\hat H] =\Or(\epsi^\infty)$ and $[\hat \pi,\hat h] =\Or(\epsi)$
imply also 
 $[\hat \pi , f(\hat H)] = \Or(\epsi^\infty)$ and  $[\hat \pi , f(\hat h)] = \Or(\epsi)$, and, $\hat \pi f(\hat H) \hat \pi =  f(\hat \pi\hat H \hat \pi) +\Or(\epsi^\infty)$ and $\hat \pi f(\hat h) \hat \pi = \hat \pi  f(\hat \pi\hat h \hat \pi)\hat \pi  +\Or(\epsi^2)$.
Modulo $\Or(\epsi^{2-n})$ we find
\begin{eqnarray*}
\Tr \left(\hat \pi f(\hat H)\, \hat a \right)  
&=& \Tr  \left(\hat\pi  f(\hat H )\, \hat\pi\, \hat a \right) =\Tr  \left(\hat\pi  f(\hat\pi \hat H \hat\pi )\, \hat\pi \,\hat a \right)  \\
&=& \Tr  \left(\hat\pi  f(\hat\pi \hat \hsc \hat\pi )\, \hat\pi \,\hat a \right) = \Tr \left(\hat\pi  f( \hat \hsc  )\, \hat\pi \,\hat a \right)   .
\end{eqnarray*}

Next note that for scalar symbols the functional calculus for pseudo-differen\-tial operators implies 
that $ f( \hat \hsc  ) - \widehat{f(   \hsc   )} = \Or(\epsi^2)$.
Hence with (\ref{traceform}) we have up to $\Or(\epsi^{2-n})$
\begin{eqnarray}\label{form1}
\Tr  \left(\hat\pi f(\hat H )\, \hat a\right)
&=&  \int \frac{\D q\D p}{(2\pi\epsi)^n}\tr \left(  f(\hsc (q,p))\, (  \pi\# a\#\pi ) (q,p)\right) \nonumber\\
&=&  \int \frac{\D q\D p}{(2\pi\epsi)^n} f(\hsc (q,p))\;\tr \left(  (  \pi\# a\#\pi ) (q,p)\right) .\nonumber
\end{eqnarray}
Using (\ref{eq:triple}) and the fact that $a$ is scalar, we have that  the expansion of 
 $A  := \pi\# a\#\pi$ starts with
$A_0  = \pi_0 a_0$ and 
\begin{eqnarray*}
A_1 &=&\pi_0 a_1\pi_0+ a_0\pi_1\pi_0+a_0\pi_0\pi_1 \\&&  - \tfrac{\I}{2}\pi_0\{a_0,\pi_0\}-  \tfrac{\I}{2}\{\pi_0, a_0\}\pi_0
-\tfrac{\I}{2}a_0\{\pi_0,\pi_0\}\,.
\end{eqnarray*}
Taking the trace we get up to $\Or(\epsi^2)$
\begin{eqnarray*}
\tr \left( \pi\# a\#\pi   \right) &=& a_0 +\epsi \left(  \tr \left( \pi_0 A_1 \pi_0   \right)+\tr \left( \pi_0^\perp A_1 \pi_0^\perp   \right)\right)   \\
&=& a_0+\epsi a_1 + \epsi  2 a_0  \tr \left( \pi_0 \pi_1 \pi_0   \right) 
\\
&=& 
a \left(1 + \I \epsi \tr \left( \pi_0 \{ \pi_0,\pi_0\}    \right)  \right)\,,
\end{eqnarray*}
where we used (\ref{eq:pb2}),  $\tr_{\hfast}(\{\pi_0,\pi_0\})=0$ and  (\ref{pi1d}).

\subsection{Egorov theorem} 
Let $\phi^t_\epsi$ be the Hamiltonian flow of $\hsc$ with respect to the symplectic form $\omega_\epsi$.
Then the Heisenberg observable
$A(t) := \E^{\I \hat H \frac{t}{\epsi}} \,\hat a \, \E^{-\I \hat H \frac{t}{\epsi}}$ 
 can be approximated by transporting  the symbol $a$  of $A(0)$  along the classical flow $\phi^t_\epsi$,
\begin{equation}\label{egorov}
  \hat\pi \left( A(t)-\widehat{a\circ \phi^t_\epsi}  \right) \hat\pi  =\mathcal{O}(\epsi^2 )\,.
\end{equation}

\noindent {\em  Proof:} With $ a(t)  :=  a\circ\phi^t_\epsi$ we need to show that 
 $  \tfrac{\df}{\df t}  \hat\pi \widehat{ a(t)}  \hat\pi = 
   \tfrac{\I}{\epsi}   [ \hat\pi \hat{H} \hat\pi, \hat\pi\widehat{ a(t)} \hat\pi  ] + \Or(\epsi^2)$.
 Let
$h_2 := \epsi^{-2}( \pi \# H \# \pi  - \pi \#  \hsc  \# \pi )$,
then
   $\pi \# h_2 \# \pi = h_2 + \Or(\epsi^\infty)$ and hence  its principal symbol 
satisfies
$(h_2)_0 = \pi_0 \,(h_2)_0\,\pi_0$.
Thus
\[
\tfrac{\I}{\epsi} \left[ \hat\pi (\hat{H}- \hat \hsc ) \hat\pi, \hat\pi\widehat{ a(t)} \hat\pi \right] 
=\I\epsi  \left[\hat h_2, \widehat{ \pi\#  a(t) \# \pi} \right]  +\Or(\epsi^\infty)   
\]
is of order $\epsi^2$ and
\begin{eqnarray*}\lefteqn{
\tfrac{\I}{\epsi}  \left[ \hat\pi \hat{H} \hat\pi, \hat\pi\widehat{ a(t)} \hat\pi \right]\;=\;
\tfrac{\I}{\epsi}  \left[ \hat\pi \hat\hsc  \hat\pi, \hat\pi\widehat{ a(t)} \hat\pi \right]  +\Or(\epsi^2) }\\
&=& \tfrac{\I}{\epsi}  \hat\pi \left[ \hat\hsc  ,\widehat{ a(t)} \right]  \hat\pi  \;-\;
\tfrac{\I}{\epsi}  \hat\pi \left(\hat\hsc  \hat\pi^\perp \widehat{ a(t)} -   \widehat{ a(t)} \hat\pi^\perp  \hat\hsc \right) \hat\pi+\Or(\epsi^2) \\
&=& \tfrac{\I}{\epsi}  \hat\pi \left[ \hat\hsc  ,\widehat{ a(t)} \right]  \hat\pi  \;+\;
\tfrac{\I}{\epsi}  \hat\pi \left[ [\hat\hsc , \hat\pi]\,,\,[ \widehat{ a(t)} ,\hat\pi] \right] \hat\pi +\Or(\epsi^2)\,.
\end{eqnarray*}
Since $\hsc $ and $ a(t)$ are scalar, (\ref{weylcommu}) implies
\begin{eqnarray*}
\tfrac{\I}{\epsi}\left[ \hat\hsc  ,\widehat{ a(t)} \right]  &=& {\rm Op}^{\rm W} ( \{\hsc  ,{ a(t)}  \}) +\Or(\epsi^2)\\
\tfrac{\I}{\epsi}\left[ \hat\hsc  ,\hat\pi \right]  &=& {\rm Op}^{\rm W} ( \{\hsc  ,\pi_0  \})+\Or(\epsi^2)\\
\tfrac{\I}{\epsi}\left[ \widehat{ a(t)}  ,\hat\pi \right]  &=& {\rm Op}^{\rm W} ( \{  a(t) ,\pi_0  \})+\Or(\epsi^2)\,,
\end{eqnarray*}
where ${\rm Op}^{\rm W} ( \cdot) := \widehat{(\cdot)}$, 
and thus, again modulo $\Or(\epsi^2)$,
\begin{eqnarray*}\lefteqn{
\tfrac{\I}{\epsi}  [ \hat\pi \hat{H} \hat\pi, \hat\pi\widehat{ a(t)} \hat\pi  ]=}\\&=&
  \hat\pi \,{\rm Op}^{\rm W} \left( \{\hsc  ,{ a(t)}  \}  -\I\epsi [  \{\hsc  ,\pi_0  \},  \{  a(t) ,\pi_0  \}]
  \right) 
   \hat\pi 
   \\
   &=& \hat\pi  \, {\rm Op}^{\rm W} \left( \{\hsc  ,{ a(t)}  \}  -\I\epsi   \tr \left( \pi_0[  \{\hsc  ,\pi_0  \},  \{  a(t) ,\pi_0  \}]\right) 
  \right) 
   \hat\pi \,.
 \end{eqnarray*}
 In order to compute $\partial_t a(t)$ recall that the Hamiltonian vector-field $X_h$ with respect to $\omega_\epsi$ is $X_h^\alpha = -(\omega_\epsi)^{\alpha\beta} \partial_\beta h$ and thus by definition of $ a(t) $ and $\Omega$ we have 
\begin{eqnarray*}
 \frac{\partial  a(t) }{\partial t} &=& \left(\hspace{-2pt} \begin{array}{c} \partial_q a\\ \partial_p a    \end{array} \hspace{-2pt}  \right)^T
 \left(\hspace{-2pt} \begin{array} {cc} - \epsi \,\Omega^{pp} &         E_n+\epsi \Omega^{pq} \\    -E_n+\epsi \Omega^{qp} &- \epsi \,\Omega^{qq} \end{array}\hspace{-2pt} \right)  \left( \hspace{-2pt}\begin{array}{c} \partial_q \hsc \\ \partial_p \hsc     \end{array}   \hspace{-2pt}\right)  \\[1mm]
 &=&  \{\hsc  ,{ a(t)}  \}  -\I\epsi\,  \tr_{\hfast}\left( \pi_0[  \{\hsc  ,\pi_0  \},  \{  a(t) ,\pi_0  \}]\right) \,.
\end{eqnarray*}

\subsection{Transport of Wigner functions}

For $\psi\in\hat\pi\Hi$ we  define the band Wigner function   as
    \[
  w_{\pi_0}^\psi :=    \left( 1-\I \epsi \tr_{\hfast}\left( \pi_0\{\pi_0,\pi_0\}\right)\right) \tr \, W^\psi   \, ,
  \]
where $W^\psi$ is the standard Wigner function. 
 Then
$ w_{\pi_0}^{\psi(t)} \;=\;  w_{\pi_0}^{\psi} \circ \phi^{-t}_\epsi + \Or(\epsi^2)$
in the sense that  
\begin{eqnarray*}
\langle \psi(t) ,\, \hat a\,\psi(t)\rangle_\Hi =   \int_{\R^{2n}}\hspace{-8pt}\df \lambda_\epsi   \,(w_{\pi_0}^{\psi}\circ \phi^{-t}_\epsi) (q,p) \,a(q,p) +\Or(\epsi^2)
\end{eqnarray*}
with $\psi(t) := \E^{-\I \hat H \frac{t}{\epsi}}\psi$ for all scalar bounded symbols $a$.\\[1mm]
\noindent {\em Proof:} Using the invariance of $\lambda_\epsi$ under the Hamiltonian flow $\phi^t_\epsi$ (Liouville's Theorem), one directly computes  
\begin{eqnarray*} \lefteqn{
  \int_{\R^{2n}}\hspace{-5pt}\df \lambda_\epsi   \, (w_{\pi_0}^{\psi}\circ \phi^{-t}_\epsi  )  \,a   
=  \int_{\R^{2n}}\hspace{-5pt}\df \lambda_\epsi   \, w_{\pi_0}^{\psi}   \,(a\circ \phi^{t}_\epsi) }  \\
&&=\; \langle \psi  ,\, \hat A(t)\,\psi \rangle_\Hi +\Or(\epsi^2) \;=\;\langle \psi(t) ,\, \hat a\,\psi(t)\rangle_\Hi+\Or(\epsi^2)\,.
\end{eqnarray*}

\subsection{Time-dependent Hamiltonians}

It is straightforward to generalize  the above statements to the case of a Hamiltonian $H(t,q,p)$ depending explicitly also on time. 
To this end one just adds the canonical pair $(t,E)$ and applies  the previous results to the symbol
$K(t,q,E,p) = E + H(t,q,p)$. Its spectral projections $\pi_0(t,q,p)$ are independent of $E$ and the classical Hamilton function is   $k(t,q,E,p) = E + h(t,q,p)$ with symplectic form $\omega = \omega_0 + \epsi\Omega(t,q,p)$, where $h $  and $\Omega$ are   computed from the instantaneous Hamiltonian $H(t,q,p)$ as before.
The equations of motion are now
\begin{eqnarray}
\dot q & = & (1   + \epsi \Omega^{pq}) \partial_ph - \epsi \Omega^{pp}\partial_q h +\epsi\Omega^{pt}\nonumber \\[-2mm] \label{tdham}\\[-2mm]
\dot p   & =& (-1  + \epsi \Omega^{qp})\partial_q h  - \epsi \Omega^{qq} \partial_p h  - \epsi \Omega^{qt}\,.\nonumber
\end{eqnarray}
On the side of quantum mechanics we have
$
(\E^{- \I  \hat K \frac{t-t_0}{\epsi}}\psi)(t) = U^\epsi(t,t_0)\psi(t_0)
$,
where $U^\epsi(t,t_0)$ is the unitary propagator generated by $\hat H(t)$. The statement of the Egorov theorem  becomes
\begin{equation}\label{egorov2} 
\hat\pi(t_0)  ( A(t)-\widehat{a\circ \phi^{t,t_0}_\epsi}  ) \hat\pi(t_0)  =\mathcal{O}(\epsi^2 )\,,
\end{equation}
where $a\circ \phi^{t,t_0}_\epsi(q,p) =  a( Q(t|t_0,q,p), P(t|t_0,q,p))$ and $  Q(t|t_0,q,p), P(t|t_0,q,p) $ is the solution to 
(\ref{tdham}) with initial data $Q(t_0|t_0,q,p) = q$ and $P(t_0|t_0,q,p) = p$.

 \subsection{The Piezo current in a magnetic field}

As an illustration consider noninteracting electrons in a slowly deformed periodic crystal  subject to a constant external magnetic field $  \vec B_0 + \epsi \vec B$, where $\vec  B_0$ has rational flux per unit cell.
The crystal is modeled by a  potential $V_\Gamma(t)$ periodic with respect to some lattice $\Gamma$. 
After a magnetic Bloch-Floquet transformation the Hamiltonian is the Weyl quantization $q\mapsto \I\epsi \nabla_p$, $p\mapsto p$, of the operator valued symbol
\begin{eqnarray*}
H_0(t,q,p) &=& \tfrac{1}{2} ( p-\I\nabla_y + \tfrac{1}{2} B_0 y +  \tfrac{1 }{2} B  q )^2 + V_\Gamma(y,t)   
\end{eqnarray*}
pointwise acting on $L^2(M_y)$, where $M$ is the fundamental domain of a magnetic super-lattice~$\tilde \Gamma$.
It is convenient to write $B_{ij} = \epsilon_{ijk} (\vec B)_k$ for the following computations. Up to unitary equivalence  $H_0(t,\kappa)$    is periodic in $\kappa = p+  \tfrac{1 }{2} B  q$ with respect to the dual lattice $\tilde\Gamma^*$ whose fundamental domain is the magnetic Brillouin zone $\mathbb{T}^*$.
 To each isolated magnetic Bloch band $e_0(t,\kappa)$ of $H_0 $ with spectral projection $\pi_0(t,\kappa)$
we  associate the classical system
$h  (t, \kappa) = e_0(t,\kappa)  + \epsi M(t,\kappa)$.
  For  $\Omega(\kappa)$ one finds 
  $\Omega^{pt} =  -\I\,\tr(\pi_0 [\nabla \pi_0 ,\partial_t \pi_0] ) $, $ \Omega^{p p }_{ij}   = -\I\,\tr(\pi_0 [\partial_{p_i}  \pi_0 ,\partial_{p_j} \pi_0] )  $, $ \Omega^{ pq}  =  \tfrac{1}{2} \Omega^{pp }B$, $ \Omega^{qq }   = - \tfrac{1}{4} B \Omega^{pp}   B$. 
  The classical equations (\ref{tdham})   thus read
 \begin{eqnarray*}
  \dot q &=&  \big(1   +\epsi\Omega^{pp}(t,\kappa)   B\big) \partial_\kappa h(t,\kappa)     +
  \epsi\Omega^{pt}(t,\kappa),\quad
  \dot \kappa = B\dot q   
 \,.
\end{eqnarray*}
Note that all classical objects are $\tilde\Gamma^*$-periodic and hence the classical phase space is really $\R^3\times \mathbb{T}^*$.
We can now approximate the current operator $J(t) := \frac{1}{\epsi}\frac{\D}{\D t} X(t)$ on the range of $\hat\pi(t_0)$ according to the Egorov theorem (\ref{egorov2}) and find that the current density   when starting in the   state $\rho(t_0) = f(\hat H(t_0))  $ is
\begin{eqnarray*}\lefteqn{ 
\lim_{\Lambda\to \R^3} \tfrac{\epsi^3}{|\Lambda|} {\rm Re}\,\Tr\left(\hat\pi(t_0) f(\hat H(t_0))   J (t)  \,\chi_\Lambda(q)\right) }\\&=&  \frac{1}{\epsi}\int_{\mathbb{T}^*}  \frac{\D \lambda_\epsi(t)}{(2\pi)^3} \left( f(h(t_0 ))\circ\phi^{t,t_0}_\epsi\right)\hspace{-2pt}(\kappa)\, \,\dot q(t ,\kappa) +\Or(\epsi )\,.
\end{eqnarray*}
With $\D \lambda_\epsi(t) = (1+\epsi  \vec\Omega (t) \cdot \vec B)\D \kappa$ and  
for a fully occupied band $f(h(t_0,\kappa ))\equiv 1$ the leading term vanishes as can be seen by partial integration. The contribution to the Piezo current density from this band is therefore
\begin{eqnarray*}
j(t) &=&    \int_{\mathbb{T}^*}  \frac{\D\kappa}{(2\pi)^3}  \left( \vec B  \,\vec \Omega (t,\kappa)     \cdot \nabla e_0(t,\kappa)     +
  \Omega^{pt}(t,\kappa)\right)\\
  &=&    \int_{\mathbb{T}^*}  \frac{\D\kappa}{(2\pi)^3}        
\,  \Omega^{pt}(t,\kappa)\,.
   \end{eqnarray*}
The last equality follows again by partial integration and the fact that $\Omega^{pp}$ is a closed $2$-form and thus  the divergence of $\vec \Omega$ with $(\vec\Omega)_i = \frac{1}{2}\epsilon_{ijk} \Omega^{pp}_{jk}$ vanishes.  From this expression it is now straightforward to derive the King-Smith and Vanderbildt formula \cite{KSV} for the orbital polarization, cf.\ \cite{Niu,PST2}. This shows that the orbital polarization is a geometric quantity that does not change under variations of the magnetic field as long as the Fermi energy lies in a gap between (magnetic) Bloch bands, see also \cite{SBT} for systems with disorder. Note, however, that the result does not follow from the modified equations of motion (\ref{tdham}) alone, but requires their correct application including the modified Liouville measure.

 \subsection{Concluding remarks } As mentioned before, the literature on adiabatic slow-fast systems is vast. Several groups \cite{LF,Niu,PST1,PST2,others} arrived independently and with different methods at equations more or less similar to (\ref{tdham}). However, their precise connection to quantum mechanical expressions was not established before for (\ref{expt}), and only shown in a special case \cite{PST1} for (\ref{egorov}).
 
 The most striking applications of the modified semiclassical model are due to Niu et al.\ \cite{Niu}.
They establish (\ref{tdham}) as the equations of motion for the center of a  Bloch  wave packet. While it is natural to conclude from this also the formulas (\ref{expt}) and (\ref{egorov2}), they give, to our knowledge,  no systematic derivation. In particular, they arrive at the  correct   Liouville measure $\lambda_\epsi$ 
 by looking for the invariant measure of (\ref{tdham}) and then postulate   (\ref{expt}). Moreover, 
 in the case of nonzero magnetic field $B_0$ the magnetic Bloch bundle defined by $\pi_0(\kappa)$ over the torus $\mathbb{T}^*$ is not trivializable, which might present  an obstruction to  patching statements about localized wave packets together. Indeed, in \cite{PST1} a rigorous derivation of (\ref{egorov}) was given for Bloch electrons with $B_0=0$, which relies heavily, as also  \cite{LW,LF2,others,LF} do,   on the possibility to chose a global non-vanishing section of the Bloch bundle. The difficulties with generalizing the approach of \cite{PST1} to magnetic Bloch bands led 
 us to the new approach presented here.

\bibliographystyle{alpha}

\end{document}